\documentstyle[12pt]{article}
\begin{document}

\begin{center}
{\large\bf ON OSCILLATORLIKE HAMILTONIANS \\ AND \\
SQUEEZING}\footnote{to be published in International Journal of
Theoretical Physics (June 2000)}\\[10mm]
\end{center}

\begin{center}
{J. BECKERS}\footnote{E-mail: Jules.Beckers@ulg.ac.be}, {N.
DEBERGH}{\footnote{E-mail:
Nathalie.Debergh@ulg.ac.be; Chercheur, Institut
Interuniversitaire des \\  Sciences Nucl\'eaires,
Bruxelles}}\\ Theoretical and Mathematical Physics, Institute of
Physics (B.5),\\ University of Li\`ege, B-4000 LIEGE 1 (Belgium)\\[5mm]
{and F.H. SZAFRANIEC}\footnote{E-mail: fhszafra@im.uj.edu.pl}\\
Instytut Matematyki, Universytet Jagiello\' nski, \\
ul. Reymonta, 4, PL-30059 Krak\' ow (Poland)\\[30mm]
\end{center}

\begin{abstract}
Generalizing a recent proposal leading to one-parameter families of
Hamiltonians and to new sets of {\it squeezed} states, we construct
larger classes of physically admissible Hamiltonians permitting new
developments in squeezing. Coherence is also discussed.\\[50mm]
\end{abstract}

\section{Introduction}
\hspace{6mm}We have recently proposed new sets of Fock states \cite{1}
which can be exploited in the contexts of coherence \cite{2} and
squeezing [3]-[5]. Due to the inclusion of a (real continuous)
parameter $\lambda$ (in the bosonic creation Heisenberg operator), the
corresponding oscillatorlike "Hamiltonians" led to stationary Schr\"
odinger problems characterized by $\lambda$-independent eigenvalues
but {\it $\lambda$-dependent eigenfunctions}, the latter ones being
particularly interesting \cite{1} for the study of new {\it squeezed}
states \cite{3}. Moreover, this approach was in a certain sense a kind
of deformation of the current one but following Wigner's point of view
\cite{6}.
\par
Let us insist strongly on the fact that we were considering \cite{1}
"squeezing" through the $\lambda$-dependent eigenfunctions of our
Schr\" odinger problems and, evidently, through the associated
meanvalues of position and momentum, their variances and (in)equalites
coming from the Heisenberg relations. This differs from Yuen's
approach \cite{3} which is based on the study of "squeezing" through
the famous two-photon coherent states of the radiation field asking
for eigenstates of the oscillator operators and not of the Hamiltonian.
\par
Here we want to generalize such developments by including a priori
more than one parameter when, simultaneously, we study new creation
{\it and} annihilation operators as well as the corresponding
oscillatorlike "Hamiltonians" appearing as physically admissible or
nonadmissible ones.
\par
The contents are then distributed as follows. In
{\it Section 2}, we recall a few relations issued from our first
approach
\cite{1}. {\it Section 3} is devoted to its generalization already
suggested. In {\it Section 4}, we apply these considerations to the
squeezing problem and find real improvements with respect to the
one-parameter previous results. Finally some conclusions and comments
are included in {\it Section 5}.

\section{A short survey of our recent proposal}
\hspace{6mm}Let us define the new (bosonic) creation Heisenberg
operator by
 \begin{equation}
a_{\lambda}^{\dagger} \equiv a^{\dagger} + \lambda I \; \; , \lambda
\in R,
\end{equation}
where $\lambda$ refers to a real continuous parameter and where
$a^{\dagger}$ is the Hermitian conjugate of the annihilation operator
$a$ satisfying altogether the expected Heisenberg commutation
relations, i.e.
\begin{equation}
[a , a_{\lambda}^{\dagger}] = I, \; [a , a] = [a_{\lambda}^{\dagger} ,
a_{\lambda}^{\dagger}] = 0.
\end{equation}
These quantum harmonic oscillatorlike considerations lead to an analog
of a (non-Hermitian) Hamiltonian of the type
\begin{equation}
H_{\lambda} = \frac{1}{2} \{a , a_{\lambda}^{\dagger}\} = \frac{1}{2}
\{a , a^{\dagger}\} + \lambda a = H_{H.O.} + \lambda a
\end{equation}
where the harmonic oscillator Hamiltonian is obviously given by
\begin{equation}
H_{H.O.} = - \frac{1}{2} \frac{d^2}{dx^2} + \frac{1}{2} x^2 \; , \;
H_{H.O.}^{\dagger} = H_{H.O.}.
\end{equation}
Moreover they ensure that
\begin{equation}
[H_{\lambda} , a] = -a, \; [H_{\lambda} , a_{\lambda}^{\dagger}] =
a_{\lambda}^{\dagger},
\end{equation}
so that (generalized) Wigner's approach \cite{6} of quantum mechanics
is still working. The energy eigenvalues and eigenfunctions have been
determined as
\begin{equation}
E_{n,\lambda} = n + \frac{1}{2} \; \; (n=0, 1, 2, ...)
\end{equation}
and
\begin{equation}
\psi_{n,\lambda} = \frac{2^{-\frac{n}{2}}
\pi^{-\frac{1}{4}}}{\sqrt{n!} \sqrt{L_n^{(0)}(-\lambda^2)}}
e^{-\frac{x^2}{2}} H_n(x+\frac{\lambda}{\sqrt{2}})
\end{equation}
where, as usual, we have chosen units such that $\omega = 1$, $\bar h
= 1$ and where $H_n$ and $L_n^{(0)}$ refer to Hermite and generalized
Laguerre polynomials \cite {7}, respectively. Let us insist on the {\it
unchanged} spectrum (6) with respect to well known oscillator
results but now with $\lambda$-modified eigenfunctions. Moreover we
have shown
\cite{1} that these new eigenfunctions (7) correspond to specific {\it
squeezed} states (\cite{3},\cite{4}). Let us recall that squeezed states
have already been experimentally detected \cite{5} being seen as
``two-photon coherent states''for the electromagnetic field. Our
new states lead to the characteristic inequality for {\it squeezing}
given by
\begin{equation}
(\Delta x)_{\lambda}^2 = 2n + \frac{1}{2} - (2\lambda^2 + 1)
\frac{L_{n-1}^{(1)}(-\lambda^2)}{L_{n}^{(0)}(-\lambda^2)} - 2\lambda^2
(\frac{L_{n}^{(1)}(-\lambda^2)}{L_{n}^{(0)}(-\lambda^2)})^2 <
\frac{1}{2}
\end{equation}
if
\begin{equation}
n = 1, 2, 3, ...\;  and \; \lambda \in R\; \backslash \;
]-r , +r[
\;, \; r \rightarrow 0 \; if \; n \rightarrow \infty.
\end{equation}

\section{A simple way to get generalized developments}
\hspace{6mm}The qualities and defects of our above approach \cite{1}
suggest the following new position of the problem:
\par
to search for (bosonic) {\it oscillatorlike} annihilation ($b$) and
creation ($b^+$) operators ensuring the following conditions
\begin{equation}
[b , b^+] = 1 \; , \; [H , b] = -b \; , \; [H , b^+] =
b^+
\end{equation}
and
\begin{equation}
H = \frac{1}{2} \{b , b^+\} = \alpha \frac{d^2}{dx^2} + \beta
(x) \frac{d}{dx} + \gamma (x),
\end{equation}
where $b$ and $b^+$ have to be general expressions of the
usual operators $a$ and $a^{\dagger}$. Let us note that we have
introduced different notations for the usual Hermitian conjugate
operator $a^{\dagger}$ of $a$ and the so-called $b^+$ associated to
$b$ with a general meaning discussed in the following.
\par
Such a set of conditions obviously {\it contains} the Heisenberg and
Wigner requirements through eqs. (10) and, moreover, restricts the
Hamiltonian to Schr\" odingerlike ones through eq. (11) where $\alpha$
is a real constant and $\beta$, $\gamma$ are arbitrary real functions
of the space variable.
\par
According to such a programme, let us introduce the generalized
operators $b$ and $b^+$ in terms of $a$ and $a^{\dagger}$ by
the following definitions
\begin{equation}
b = (1 + c_1) a + c_2 a^{\dagger} + c_3
\end{equation}
and
\begin{equation}
b^+ = c_4 a + (1 + c_5) a^{\dagger} + c_6,
\end{equation}
where $c_1, c_2, ..., c_6$ are arbitrary (real) parameters and where
the current harmonic oscillator context has been included by equating
all the parameters to zero while the definition (1) is also
incorporated in equating only $c_6$ with the $\lambda$-parameter, all
the other $c$'s being identically zero.
\par
At this stage let us point out that the generalized operators (12) and
(13) are intimately connected with the construction of the so-called
"two-photon coherent states" due to Yuen \cite{3} but here from
inhomogeneous linear canonical transformations which could be
summarized by the following form
$$
\left(
\begin{array}{ccc}
b \\
b^+
\end{array}
\right)
=
\left(
\begin{array}{ccc}
1+c_1 & c_2 \\
c_4 & 1+c_5
\end{array}
\right)
\left(
\begin{array}{ccc}
a \\
a^{\dagger}
\end{array}
\right)
+
\left(
\begin{array}{ccc}
c_3\\
c_6
\end{array}
\right)
$$
where the matrix has to be invertible so that, for example,
$$
(1+c_1) (1+ c_5) - c_2c_4 = 1.
$$
This leads to
\begin{equation}
c_1 + c_5 + c_1c_5 - c_2c_4 = 0
\end{equation}
which is the constraint between the $c$'s issued from (10) and
leaving, in fact, only five independent parameters in the whole
discussion.
\par
 By taking care of the definitions (12) and (13) in the Hamiltonian (11)
and by remembering that
\begin{equation}
a = \frac{1}{\sqrt{2}} (\frac{d}{dx} + x) \; , \; a^{\dagger} =
\frac{1}{\sqrt{2}} (-\frac{d}{dx} + x),
\end{equation}
the possible Hamiltonians are then found on the following form
\begin{equation}
H = A \frac{d^2}{dx^2} + (B x + C) \frac{d}{dx} + Dx^2 + Ex + F
\end{equation}
transferring the parametrization on the six parameters $A, B, ..., F$
given by
\begin{eqnarray}
&&A = - \frac{1}{2} - c_2c_4 + \frac{1}{2}c_4(1+c_1) +
\frac{1}{2}c_2(1+c_5), \nonumber \\
&&B = c_4(1+c_1) - c_2(1+c_5), \nonumber \\
&&C = \frac{1}{\sqrt{2}} [c_6 (c_1-c_2+1) + c_3 (c_4-c_5-1)],\nonumber
\\ &&D =  \frac{1}{2} + c_2c_4 + \frac{1}{2}c_4(1+c_1) +
\frac{1}{2}c_2(1+c_5),  \\
&&E = \frac{1}{\sqrt{2}} [c_6 (c_1+c_2+1) + c_3 (c_4+c_5+1)],\nonumber
\\ &&F = \frac{1}{2} c_4(1+c_1) - \frac{1}{2}c_2(1+c_5) + c_3c_6.
\nonumber
\end{eqnarray}
These developments compared to the previous ones \cite{1}
clearly appear as a generalization; moreover it permits an
interesting discussion at the level of physically admissible
Hamiltonians as well as at the level of coherence and (or) squeezing,
once we have solved the eigenvalue and eigenfunction problems
associated with such Hamiltonians.
\par
In terms of the new parameters, let us point out again that the
current harmonic oscillator Hamiltonian (4) corresponds to
\begin{equation}
A = -D = - \frac{1}{2}\; , \; B = C = E = F = 0,
\end{equation}
while our previous deformation (1) leading to the Hamiltonian (3) is
given by
\begin{equation}
A = -D = - \frac{1}{2}\; , \; B = F = 0\; , \; C = E =
\frac{\lambda}{\sqrt{2}}.
\end{equation}
\par
With the Hamiltonian (16) and the relations (17), the stationary
Schr\" o-\\ dinger problem can now be solved by conventional quantum
mechanical methods \cite{8}. It leads to the general answer
\begin{equation}
E_n =
F - \frac{B}{2} - \frac{C^2}{4A} -\frac{A}{p^2} (2n + 1) + q^2 (D -
\frac{B^2}{4A}) + q (E - \frac{BC}{2A})
\end{equation}
while the corresponding eigenfunctions take the form
\begin{equation}
\psi_n(x) = exp[ -\frac{B}{4A} x^2 - \frac{C}{2A} x] exp[ -
\frac{x^2}{2p^2} + \frac{qx}{p^2} - \frac{q^2}{2p^2}]  H_n
(\frac{x-q}{p}),
\end{equation}
where $p$ and $q$ enter the necessary change of variable
\begin{equation}
x = py + q,\; (p \neq 0).
\end{equation}
Let us mention the two constraints
\begin{equation}
\frac{p^4}{A} (D - \frac{B^2}{4A}) = -1
\end{equation}
and
\begin{equation}
2q  (D - \frac{B^2}{4A}) + E - \frac{BC}{2A} = 0,
\end{equation}
issued from these calculations.Together with eqs. (17), these relations
(23), (24) fix the parameters $p$ and $q$ of our change of variable
(22) to the unique values:
\begin{equation}
p^2 = -2A \; , \; q = 2EA - BC
\end{equation}
in order to get in particular a positive spectrum. By requiring to
deal with square integrable eigenfunctions, we finally have to ask for
\begin{equation}
A < 0 \;\; \;  and \;\; \; B < 1
\end{equation}
compatible with the specific cases (18) and (19). We thus get (up to a
normalization factor $N_n$) the solutions (21) as given by
\begin{eqnarray}
&&\psi_n(x) = N_n exp[ \frac{1-B}{4A} x^2 ] exp[
(\frac{(B-1)C}{2A}- E)x] exp[\frac{(2EA-BC)^2}{4A}]\nonumber \\ && H_n
(\frac{1}{\sqrt{-2A}}(x - 2EA + BC)).
\end{eqnarray}
Moreover we obtain the remarkable result of an {\it unchanged} real
spectrum
\begin{equation}
E_n = n + \frac{1}{2} \; , \; n = 0, 1, 2, ...,
\end{equation}
inside this general context as it was already the case in our first
study (see (6)). Let us here insist on this real character without
having required the selfadjointness of the Hamiltonian (16), a similar
property to the one which has recently been quoted by Bender and
Boettcher \cite{9} although we have not required any specific discrete
symmetries. Nevertheless, we have also noticed that a {\it necessary}
and {\it sufficient} condition ensuring that
$H^{\dagger} = H$ is simply
\begin{equation}
B = C = 0
\end{equation}
leading to a large class of {\it physically} admissible Hamiltonians.
\par
As a last remark in this Section, let us point out that such
eigenfunctions $\psi_n(x)$ like (27) are once again associated with
Fock states - let us call them $\mid n >_c$ referring to the
$c$-parametrization included in eqs.(12) and (13) - and it is
interesting to quote the action of $b$ and $b^{\dagger}$ on such
states. We obtain
\begin{equation}
b \mid n >_c = \frac{n}{\sqrt{-A}} \frac{N_n}{N_{n-1}} (1+c_1-c_2)
\mid n-1 >_c
\end{equation}
and
\begin{equation}
b^+ \mid n >_c = \frac{1}{2\sqrt{-A}} \frac{N_n}{N_{n+1}}
(1+c_5-c_4)
\mid n+1 >_c
\end{equation}
and point out that
\begin{equation}
b b^+ \mid n >_c = (n+1) \mid n >_c \; , \;  b^+b \mid
n >_c = n \mid n >_c
\end{equation}
so that the conditions (10) and (11) are obviously satisfied, ensuring
in particular that
\begin{equation}
\{b , b^+\}\mid n >_c  = 2H \mid n >_c  = (2n+1)\mid n >_c.
\end{equation}

\section{On implications in squeezing}
\hspace{8mm} Let us, {\it first}, extract new information by considering
the lowest energy eigenvalue $E_0$ of the spectrum and its associated
eigenfunction $\psi_0(x)$. Then, as a {\it second} step, let us come
back very briefly on known cases corresponding to the Hamiltonian (16)
with the conditions (18) or (19). Finally, let us consider {\it
thirdly} new parametrizations exploiting the results obtained in
Section 3 mainly with a view of interesting improvements in squeezing.
\subsection{From the lowest eigenvalue of the spectrum}
\hspace{8mm} Due to the fundamental and specific role played by the
lowest energy eigenvalue $E_0$, let us study coherence and squeezing
through the eigenfunction $\psi_0(x) \equiv (27)$ which takes the
explicit form
\begin{equation}
\psi_0(x) = N_0 exp[ - \frac{1}{2} \alpha x^2 - \beta x - \frac{1}{2}
\gamma]\; , \; N_0 = (\frac{\alpha}{\pi})^{\frac{1}{4}}
exp[\frac{\alpha \gamma - \beta^2}{2\alpha}]
\end{equation}
in order to ensure that
\begin{equation}
\int_{-\infty}^{+\infty} \psi_0^2(x) dx = 1
\end{equation}
where, for brevity, we have introduced the notations
\begin{equation}
\alpha= \frac{B-1}{2A} \; , \; \beta = E - \frac{(B-1)C}{2A} \; , \;
\gamma = - \frac{1}{2A} (2EA - BC)^2.
\end{equation}
\par
In such a $n=0$-context, the meanvalues and consequences are readily
obtained as follows:
\begin{eqnarray}
&&\langle x \rangle _0 = \frac{2EA+C}{1-B}\; , \; \langle x^2 \rangle
_0 = \frac{A}{B-1} + (\frac{2EA}{B-1} - C)^2, \nonumber \\
&&\langle p \rangle _0 = 0 \; , \; \langle p^2 \rangle _0 =
\frac{B-1}{4A}\; ,
\end{eqnarray}
so that we get
\begin{equation}
(\Delta x)_0^2 = \frac{A}{B-1} \; , \; (\Delta p)_0^2 = \frac{B-1}{4A}.
\end{equation}
These results ensure {\it coherence} due to the Heisenberg relation
\begin{equation}
(\Delta x)_0(\Delta p)_0 = \frac{1}{2}
\end{equation}
and {\it squeezing} on the x-variable
\begin{equation}
(\Delta x)_0^2 < \frac{1}{2} \; \; \; \;  iff \; \; \; \; B < 2A+1
\end{equation}
{\it or} on the p-variable
\begin{equation}
(\Delta p)_0^2 < \frac{1}{2} \; \; \; \;  iff  \; \; \; \; B > 2A+1.
\end{equation}
Such inequalities on $A$ and $B$ only will suggest our future
parametrizations in Sections 4.2 and 4.3. In fact, let us immediately
inform the reader that we plan to priviledge the discussion on the
x-variable so that eq. (40) will play the main role.
\subsection{From known cases}
\hspace{8mm} (i) The {\it harmonic oscillator} context characterized by
the condition (18) is obviously well known as far as coherence and
squeezing are visited (\cite{2}-\cite{5}). As already mentioned, this
case is contained in our study but we learn only that it corresponds
to all $c$'s equal to zero in eqs. (12) and (13), it generates a
selfadjoint Hamiltonian (4) and deals with hermitian conjugated
operators $b \equiv a$ and $b^{\dagger} \equiv a^{\dagger}$ verifying
the condition
\begin{equation}
(b^{\dagger})^{\dagger} = b.
\end{equation}
\par
(ii) The {\it deformed} context characterized by the condition (19)
has already been discussed in \cite{1}: it breaks down the condition
(42) {\it and} the selfadjointness of the Hamiltonian (3) so that
physical connections are here questionable although they correspond to
{\it real} spectra and to new possibilities of squeezing for $n \neq
0$ \cite{1}. Let us point out that the conditions (29) are obviously
in contradiction with eqs. (19) and that, for $n=0$, the inequalities
(40) cannot be satisfied.
\subsection{To new contexts}
\hspace{8mm}  By keeping the conditions (29) in order to maintain the
selfadjointness of the Hamiltonians, we can also require the condition
(42). The latter leads to very simple demands of the types:
\begin{equation}
c_1 = c_5 \; , \; c_2 = c_4 \; , \; c_3 = c_6
\end{equation}
so that we then get families of {\it physically admissible
Hamiltonians} which can be further exploited.\\
\vspace{5mm}
\par
(i) Within such conditions, let us go to a one-parameter
$\lambda$-deformation with, for example, the values
\begin{equation}
c_1 = c_5 = \frac{2}{3} \; , \; c_2 = c_4 = \frac{4}{3} \; , \; c_3 =
c_6 = \lambda.
\end{equation}
Such a case corresponds to the parametrization (17) given by
\begin{equation}
A = - \frac{1}{18}\; , \; C = \frac{9}{2}\; , \; E = 3 \sqrt{2}
\lambda \; , \; F = \lambda^2\; , \; B=C=0
\end{equation}
and the eigenvalues and eigenfunctions problem can be completely
solved. We evidently get the spectrum (28) and the eigenfunctions (27)
take the final form
\begin{equation}
\psi_n(x) = N_n exp[- \frac{9}{2} x^2 - \frac{6}{\sqrt{2}} \lambda x -
\lambda^2] \;  H_n(3x + \sqrt{2} \lambda)
\end{equation}
where the normalization factor is found on the following form
\begin{equation}
N_n = \frac{\sqrt{3} \pi^{-\frac{1}{4}}
2^{-\frac{n}{2}}}{\sqrt{n!}}.
\end{equation}
Meanvalues and Heisenberg constraints can then be evaluated and we get
\begin{equation}
\langle x \rangle _{\lambda} = -\frac{\sqrt{2}}{3} \lambda\; , \;
\langle x^2
\rangle _{\lambda} = \frac{1}{9}(2 \lambda^2 + n + \frac{1}{2})
\end{equation}
and
\begin{equation}
\langle p \rangle _{\lambda} = 0 \; , \; \langle p^2 \rangle
_{\lambda} = 9 (n + \frac{1}{2}),
\end{equation}
so that
\begin{equation}
(\Delta x)_{\lambda}^2 = \frac{1}{9} (n + \frac{1}{2}) \; , \; (\Delta
p)_{\lambda}^2 = 9 (n + \frac{1}{2})
\end{equation}
leading to
\begin{equation}
(\Delta x)_{\lambda}(\Delta p)_{\lambda} = n + \frac{1}{2}.
\end{equation}
This result is analogous to the one of the {\it undeformed} case but,
here, it permits, moreover, squeezing (but on x {\it only}) due to the
relations (40) and (50). In fact, such a squeezing can only take
place for $n= 0, 1, 2, 3.$\\
\vspace{5mm}
\par
(ii) A final improvement of this example consists in the possible
increase of such $n$-values permitting the squeezing and maintaining
the nice property (42) and the selfadjointness of $H$. This can be
realized through the new $\lambda$-deformation ($\lambda > 0$)
characterized by the values
\begin{equation}
c_1 = c_5 = \frac{(\sqrt{\lambda} - 1)^2}{2 \sqrt{\lambda}} \; , \;
c_2 = c_4 = \frac{\lambda - 1}{2 \sqrt{\lambda}} \; , \; c_3 = c_6 = 0,
\end{equation}
leading to the relations (17) given now on the form
\begin{equation}
A = - \frac{1}{2 \lambda}\; , \; D = \frac{\lambda}{2} \; , \; B = C =
E = F = 0.
\end{equation}
satisfying once again the inequalities (40) when $n=0$.
\par
Here the eigenfunctions are found as
\begin{equation}
\psi_n(x) = N_n exp[- \frac{\lambda}{2} x^2] \; H_n (\sqrt{\lambda} x)
\; , \;  N_n = \frac{\lambda^{\frac{1}{4}} \pi^{-\frac{1}{4}}
2^{\-\frac{n}{2}}}{\sqrt{n!}}
\end{equation}
and we get in correspondence with eqs. (50)
\begin{equation}
(\Delta x)^2_{\lambda} = \frac{1}{\lambda} (n + \frac{1}{2}) \; , \;
(\Delta p)^2_{\lambda} = \lambda (n + \frac{1}{2}).
\end{equation}
We thus notice once more the validity of eq. (51) ensuring {\it
coherence} for the particular value $n=0$ only but {\it squeezing} (in
the $x$-coordinate) for all the values $n$ satisfying the following
inequality
\begin{equation}
\lambda > 2n+1 > 0.
\end{equation}
A further interesting property of the above eigenfunctions (54) (and
evidently (46)) is that, due to the characteristics of Hermite
polynomials \cite{7}, these solutions are not only normalized but are
also {\it orthogonal} as it can be easily established.
\par
If physical applications require a fixed {\it finite} set of levels in
the energy spectrum, we can always choose, due to the inequality (56),
our $\lambda$-parameter in order to guarantee the squeezing up to this
$n$-value.\\
\vspace{5mm}
\par
(iii) As a last context, let us relax the condition (42) and the
selfadjointness of the Hamiltonian. This corresponds to an extension
of the context discussed in \cite{1} and recalled here in the
subsection (4.2.ii). We can choose, for example,
\begin{equation}
b = a + \lambda a^{\dagger} \; , \; b^+ = a^{\dagger}
\end{equation}
corresponding to all the null parameters $c$ except $c_2 = \lambda$ or
to
\begin{equation}
A = \frac{1}{2} (\lambda -1)\; , \; B = - \lambda \; , \; C = E = 0 \;
, \; D = \frac{1}{2} (\lambda +1) \; , \; F = - \frac{\lambda}{2}
\end{equation}
ensuring squeezing on $x$ in the $n=0$-case if $1 > \lambda > 0$. With
the spectrum (28), the associated eigenfunctions here take the form
\begin{equation}
\psi_n(x) = N_n exp[-\frac{1}{2} (\frac{1 + \lambda}{1 - \lambda})
x^2] \; H_n (\frac{x}{\sqrt{1-\lambda}}).
\end{equation}
They are normalizable with
\begin{equation}
N_n = \frac{\pi^{-\frac{1}{4}}}{n!}
(\frac{1+\lambda}{1-\lambda})^{\frac{1}{4}} (1+\lambda)^{\frac{n}{2}}
F_n^{-\frac{1}{2}} (\lambda)
\end{equation}
but not orthogonal. Depending on the {\it even} or {\it odd} character
of $n$ the functions $F_n(\lambda)$ are respectively given by
\begin{equation}
F_n (\lambda) = \sum_{l=0}^{\frac{n}{2}} \frac{2^{2l}
\lambda^{n-2l}}{(2l)![(\frac{n}{2} - l)!]^2} \; \; , \; (n \; \;
even),
\end{equation}
or
\begin{equation}
F_n (\lambda) = \sum_{l=0}^{\frac{n-1}{2}} \frac{2^{2l+1}
\lambda^{n-1-2l}}{(2l+1)![(\frac{n-1}{2} - l)!]^2} \; \; , \; (n
\; \; odd).
\end{equation}
These functions enter the evaluation of meanvalues and Heisenberg
constraints for each $n$-value. Specific values are of interest in
order to learn the general behaviour of the corresponding meanvalues
and their consequences but these are only exercises. Let us just point
out that, for $n=0$, we get
\begin{eqnarray}
&&\langle x \rangle _{\lambda} = 0\; , \;
\langle x^2
\rangle _{\lambda} = \frac{1}{2} (\frac{1-\lambda}{1+\lambda}) = (\Delta
x)^2_{\lambda} \nonumber \\
&&\langle p \rangle _{\lambda} = 0 \; , \;
\langle p^2
\rangle _{\lambda} = \frac{1}{2} (\frac{1+\lambda}{1-\lambda}) = (\Delta
p)^2_{\lambda}
\end{eqnarray}
giving us {\it coherence} due to
\begin{equation}
(\Delta x)^2_{\lambda} (\Delta p)^2_{\lambda} = \frac{1}{4} \; \; ,
\; \; \; \forall \lambda,
\end{equation}
while {\it squeezing} requires parametrizations according to
\begin{equation}
1 > \lambda > 0 \; \; \; or \; \; \; -1 < \lambda < 0
\end{equation}
in the $x$- {\it or} $p$- context respectively. Coherence is then lost
if $n \neq 0$ but squeezing can be installed when specific refined
inequalities of the type (65) are valid. The upper and lower bounds on
these $\lambda$-values can be determined by entering the
results (59)-(62) depending on the $n$-values we are considering.
\section{Some further conclusions and comments}
\hspace{8mm}
Among the above results, let us point out those obtained more
particularly in the subsection (4.3.ii) leading to an attracting class
of one-parameter selfadjoint Hamiltonians
\begin{equation}
H_{\lambda} = \frac{1}{2} \{b_{\lambda} , b_{\lambda}^{\dagger} \},
\end{equation}
with
\begin{equation}
b_{\lambda} = (1 + \frac{(\sqrt{\lambda} - 1)^2}{2\sqrt{\lambda}}) a +
\frac{\lambda - 1}{2\sqrt{\lambda}} a^{\dagger}\; , \;
b_{\lambda}^{\dagger} = b^+ \; , \;  (b_{\lambda}^{\dagger})^{\dagger} =
b_{\lambda},
\end{equation}
characterized by a deformation parameter $\lambda > 0$ and
corresponding to the current harmonic oscillator case when $\lambda =
1$. Appearing nearly as a trivial result, this family opens possible
new studies of squeezing through energy eigenfunctions (54) which are
not only normalizable but also orthogonal among themselves. The
possible choice $\lambda > 2n+1$ with a fixed set of energy
eigenvalues given by the usual spectrum (28) is maybe an interesting
connection with possible experimental realizations for oscillatorlike
systems in order to test and to realize the associated squeezed states.
\par
One further comment is the possible exploitation of our generalized
operators $b$ and $b^+$ by studying more than one (real) parameter in
the definitions (12) and (13) as just noticed elsewhere \cite{11}.
\par
Another point which has to be recalled is that the motivations of
deforming our (annihilation and creation) oscillatorlike operators (as
realized in eqs. (12) and (13)) were intimately connected with a
specific mathematical property called ``subnormality of operators''
\cite{10}, a property already exploited in our previous letter \cite{1}.
\par
As a final comment, let us also recall that our developments could
evidently be extended to the fermionic sector as already specified in
our first approach \cite{1}. The generalizations (12) and (13) can be
realized on {\it fermionic} annihilation and creation operators and
their consequences can then be deduced. Then the superposition of
these bosonic and fermionic contexts could be considered in order to
go towards supersymmetric developments [12] by including
simultaneously specific physical as well as mathematical properties.

\end{document}